\documentclass[aps,prl,twocolumn]{revtex4-1}

\usepackage{graphicx}
\usepackage{dcolumn}
\usepackage{bm}
\usepackage{braket}
\usepackage{dsfont}
\usepackage{amsmath}
\usepackage{subfigure}
\usepackage{extarrows}
\usepackage{amssymb}
\usepackage{esint}
\usepackage{qcircuit}
\usepackage{tikz}
\usetikzlibrary{trees,arrows}
\usepackage{xcolor}
\usepackage{hyperref}

\begin{document}

\newcommand{\uvec}[1]{\boldsymbol{\hat{\textbf{#1}}}}
\newcommand\logeq{\mathrel{\vcentcolon\Leftrightarrow}}
\newcommand{\redmark}[1] {\color{red}\textbf{#1}\color{black}\normalsize}
\newcommand{\bluemark}[1] {\color{blue}\textbf{#1}\color{black}\normalsize}
\newcommand{\brownmark}[1] {\color{purple}\textbf{#1}\color{black}\normalsize}

\title{Berry Phase Estimation in Gate-Based Adiabatic Quantum Simulation}

\author{Bruno Murta}
\email{bruno.murta@inl.int}
\author{G. Catarina}
\author{J. Fern\'{a}ndez-Rossier}
\altaffiliation[Also at ]{Departamento de F\'isica Aplicada, 03690 San Vicente del Raspeig,
 Universidad de  Alicante, Spain}
\affiliation{%
QuantaLab,  International Iberian Nanotechnology Laboratory (INL)\\
 Avenida Mestre Jos\'{e} Veiga \\
 4715-330 Braga, Portugal\\\
}%

\date{\today}

\begin{abstract}

Gate-based quantum computers can in principle simulate the adiabatic dynamics of a large class of Hamiltonians. Here we consider the cyclic adiabatic evolution of a parameter in the Hamiltonian. We propose a quantum algorithm to estimate the Berry phase and use it to classify the topological order of both single-particle and interacting models, highlighting the differences between the two. This algorithm is immediately extensible to any interacting topological system. Our results evidence the potential of near-term quantum hardware for the topological classification of quantum matter.

\end{abstract}

\pacs{Valid PACS appear here}
\maketitle

Fault-tolerant universal quantum computers are expected to efficiently simulate the unitary evolution of large classes of quantum Hamiltonians \cite{abramslloyd1997,abramslloyd1999,troyer2015}, including those relevant for condensed matter \cite{Zhang2012}, quantum chemistry \cite{Lanyon}, and sub-atomic physics \cite{dumitrescu2018}. In particular, they will help to address the exponential wall problem \cite{Kohnnobel1999} faced in the simulation of quantum many-body phenomena.   

Algorithms for the preparation of complicated quantum states are required in most digital quantum simulation (DQS)  strategies. In some instances, such as hybrid variational methods \cite{kandala2017} and phase estimation \cite{aspuru-guzik2005}, the preparation of approximate quantum states is a valid approach, as long as the overlap with the target exact state is large enough. However, this overlap is expected to become exponentially small as the number of degrees of freedom increases \cite{anderson1967}. A solution to this problem is parametric adiabatic evolution via DQS \cite{barends2016}. Starting from a Hamiltonian for which the ground state can be easily obtained, the extra terms are added slowly, and, by virtue of the adiabatic theorem \cite{BornFockAdiabatic}, the quantum state of the system stays in the ground state of the new Hamiltonian. 

A central concept in the theory of adiabatic parametric evolution is the Berry phase \cite{Berry}. As a Hamiltonian is cycled adiabatically around a closed path in a parameter space, the wave function acquires a geometric phase \cite{Berry} in addition to the dynamical phase. 
The Berry phase plays a crucial role in several domains of quantum theory \cite{wilczek1989}, including our understanding of  electronic properties of molecules \cite{resta2000}, nanomagnets \cite{von1992,wernsdorfer1999}, solids \cite{xiao2009,vanderbilt2018} and the topological theory of quantum  matter \cite{hasankane2010,qizhang2011}. Specifically, the Berry phase can be used as a quantized index for the topological classification of different classes of Hamiltonians, including one-dimensional symmetry-protected topological insulators \cite{delplace2011,cao17,velasco17}, gapped spin liquids \cite{Hatsugai2006} and interacting fermion models \cite{le2019}.  


As one of the main platforms for quantum simulation, superconducting qubits have been used to explore topological states. Quantum algorithms to measure single-particle topological invariants, one based on quantum walks \cite{Flurin2017} and another for finite temperatures \cite{Viyuela2016}, have been recently proposed. A more general method was used to probe topological transitions in both single-qubit \cite{Schroer2014} and coupled two-qubit \cite{Roushan2014} systems. This involved the measurement of deflections from the adiabatic path to obtain the local Berry curvature \cite{Gritsev2012}, which was then integrated to obtain the Berry phase. 
  
In this Letter we propose a quantum algorithm that yields the Berry phase without requiring the explicit integration of the Berry curvature. Our algorithm combines phase estimation and gate-based  simulation of adiabatic quantum evolution to obtain  the Berry phase, as opposed to  the so-called adiabatic quantum computing \cite{albashlidar2018}. 
This algorithm can be applied to a wide class of Hamiltonians in a parameter space. In particular, we show how it can be used for the topological classification of model Hamiltonians with gapped ground states, working out the cases of both the paradigmatic Su-Schrieffer-Heeger (SSH) Hamiltonian \cite{ssh1979} for independent fermions and the  dimerized Heisenberg $S=1/2$ spin chain \cite{fields1979}. 

The formal statement of the problem addressed here  is the following. Given a family of Hamiltonians ${\cal H}(\rho)$  obtained from variations of a parameter $\rho$, we focus on the case where, for every $\rho$,  ${\cal H}(\rho)$ has a non-degenerate ground state $|\Psi_G(\rho)\rangle$ with energy $E_G(\rho)$.   At $t=0$, $\rho(t=0) \equiv \rho_0$ and the system is prepared in its ground state $|\Psi_G(\rho_0)\rangle$. The system evolves in time as $\rho$ changes slowly enough to ensure that it remains in the ground state $|\Psi_G(\rho)\rangle$ per the adiabatic theorem \cite{BornFockAdiabatic}. After a time $T$, $\rho = \rho_T$ and ${\cal H}(\rho_T)={\cal H}(\rho_0)$.  Without loss of generality $\rho$ can be considered to be an angle that varies between $0$ and $2\pi$ and ${\cal H}$ to depend on $\rho$ via periodic functions. The  parametric evolution can thus be visualized as a loop in the unit circle generated by $\rho \in [0,2\pi)$.
 
The quantum state at $t=T$ adopts the form
 \begin{equation}
 |\Psi_G(2\pi )\rangle=e^{-i\theta_D} e^{i \theta_B}|\Psi_G(0)\rangle,
 \end{equation}
\noindent where $\theta_D = \frac{1}{\hbar} \int_0^T E_G(\rho(t)) \; dt$  is the dynamical phase and 
\begin{equation}
\theta_B=-i \int_0^{2\pi} \langle \Psi_G(\rho)|\frac{\partial \Psi_G(\rho)}{\partial \rho}\rangle d\rho
\label{B}
\end{equation}

\noindent is the Berry phase. Importantly, the Berry phase $\theta_B$ is symmetric under time reversal, whereas the dynamical phase $\theta_D$ is anti-symmetric. Our goal is to carry out a gate-based quantum simulation of the adiabatic loop to determine $\theta_B$. This is accomplished by a combination of quantum phase estimation \cite{abramslloyd1999} and gate-based quantum simulation of the adiabatic evolution. The proposed  quantum circuits are shown in Fig. 1. 
  
\begin{figure}
\includegraphics[width=0.95\linewidth]{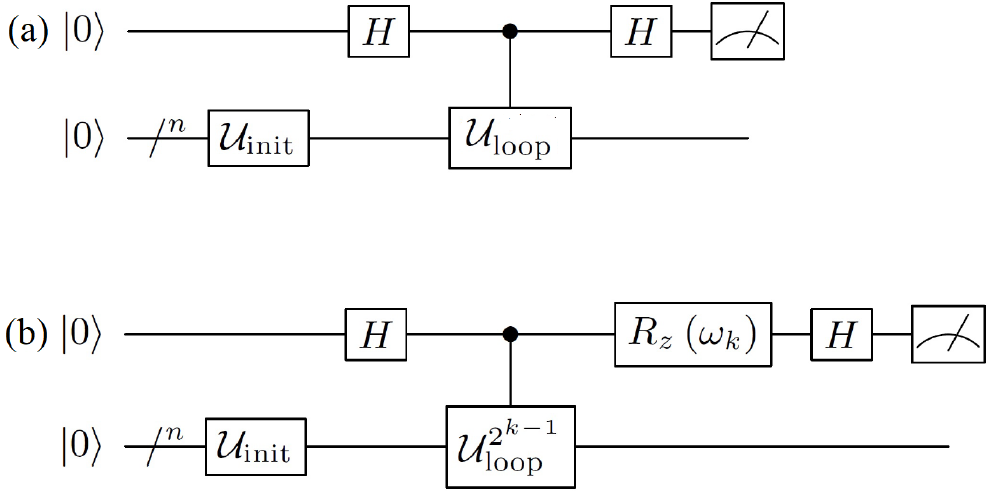}
\caption{Quantum circuits to measure Berry phase. $\mathcal{U}_{\rm init}$ represents subcircuit that initializes n-qubit register in eigenstate of $\mathcal{H}$.   $\mathcal{U}_{\rm loop}$ implements the quantum simulation of the adiabatic loop. $H$ are Hadamard gates. (a) Hadamard test scheme \cite{cleve98}. (b) Iterative Phase Estimation (IPE) scheme \cite{Dobsicek}. Quantum circuit shows k\textsuperscript{th} iteration. The $R_z(\omega_k)$ gate, where $\omega_k = -2\pi0.0...\gamma_{k+1}...\gamma_R$ and $R$ is the total number of iterations, serves to remove the contribution to the phase from the previously measured bits.}
\label{fig1}
\end{figure}

We first discuss the circuit shown in Fig. 1(a), which represents the standard interferometric phase estimation circuit \cite{cleve98}. An ancilla qubit reads out the Berry phase, and a n-qubit register stores the quantum state that undergoes the evolution.  The initialization subroutine, denoted by $\mathcal{U}_{\rm init}$, accomplishes $\mathcal{U}_{\rm init}|0\rangle_n= |\psi_G\rangle$. 

The crux of the matter lies on the second stage, which carries out the controlled adiabatic evolution ${\cal U}_{\rm loop}|\psi_G\rangle= e^{i\phi} |\psi_G\rangle$, where $\phi$ depends both on the dynamical and the Berry phases. Performing adiabatic evolution for the state initialization prior to the phase estimation scheme has been previously explored, but instead we introduce the adiabatic evolution within the phase estimation itself. The combination of the Hadamard gate on the ancilla and the controlled operation kick the $\phi$ phase onto the top register. As a result, in  the last stage of the process,  the probability of the ancilla being measured as 0 is $P_0= \cos^2\Big(\frac{\phi}{2}\Big)$ (see supp. mat. and \cite{cleve98}). 

In this work we also use the so-called iterative phase estimation  algorithm (IPEA) \cite{Dobsicek}, the output of which is the phase itself expressed as an R-bit binary fraction of the form $\frac{\phi}{2\pi}= \sum_{k=1,R} \frac{\phi_k}{2^k}$, where $R$ is the number of iterations.  The binary digits $\phi_k$ are obtained by repeatedly applying the circuit at the bottom of Fig. 1. $R_z(\omega_k)$ uses the results of the previous steps to gauge away the corresponding phase in the ancilla qubit, thus ensuring that at the k\textsuperscript{th} iteration the circuit yields the digit $\phi_{R-k}$. 

A crucial element of the generality of our algorithm lies in the structure of ${\cal U}_{\rm loop}$. Specifically, we take:
\begin{equation}
{\cal U}_{\rm loop} = U_{\circlearrowleft} (0,T/2) U_{\circlearrowleft}(T/2,0),
\label{time_reversal}
\end{equation}
\noindent where the first (second) argument stands for forward evolution in time from $t=0$ to $t=T/2$ (backward evolution in time from $t=T/2$ to $t=0$) and the subindex $\circlearrowleft$ denotes counter-clockwise evolution in $\rho$-space. In words, $\rho$  always changes counter-clockwise from $0$ to $2\pi$, but time evolves forward through the first half of the single $\rho$-loop and then backward. This allows to cancel the dynamical phase whilst keeping the Berry phase, yielding $\phi=\theta_B$. The sole drawback of this approach is that it is only valid for time-reversal-symmetric Hamiltonians. 

The implementation of the adiabatic evolution quantum subroutine $ {\cal U}_{\rm loop}$  in gate-based quantum computers, such as the IBM Q Experience devices, is accomplished by breaking down the evolution  in $N$ steps of duration $\delta t$ during which $\rho$ stays constant: 
\begin{equation}
\mathcal{U}_{\rm loop} = \prod_{j = 1}^N \delta \mathcal{U}(\rho_j).
\label{Uprod}
\end{equation}
\noindent $\delta \mathcal{U}(\rho_j) = \exp(-i \: \mathcal{H}(\rho_j) \: \delta t / \hbar)$ stands for the unitary propagator element due to the Hamiltonian ${\cal H}(\rho_j)$, keeping $\rho$ constant. The choice of both $T$ and $N$  is determined by two competing factors. On the one hand, the adiabatic condition requires that $\frac{\delta \rho}{\delta t}=\frac{2\pi}{T}$ is small, which imposes large enough N (cf. $\delta \rho= \frac{2\pi}{N}$) and T.  On the other hand, the number of gates in the quantum simulation algorithm increases with both $N$ and $T$, which is an issue given the limitations of current quantum hardware.  
\begin{figure}
\includegraphics[width=0.95\linewidth]{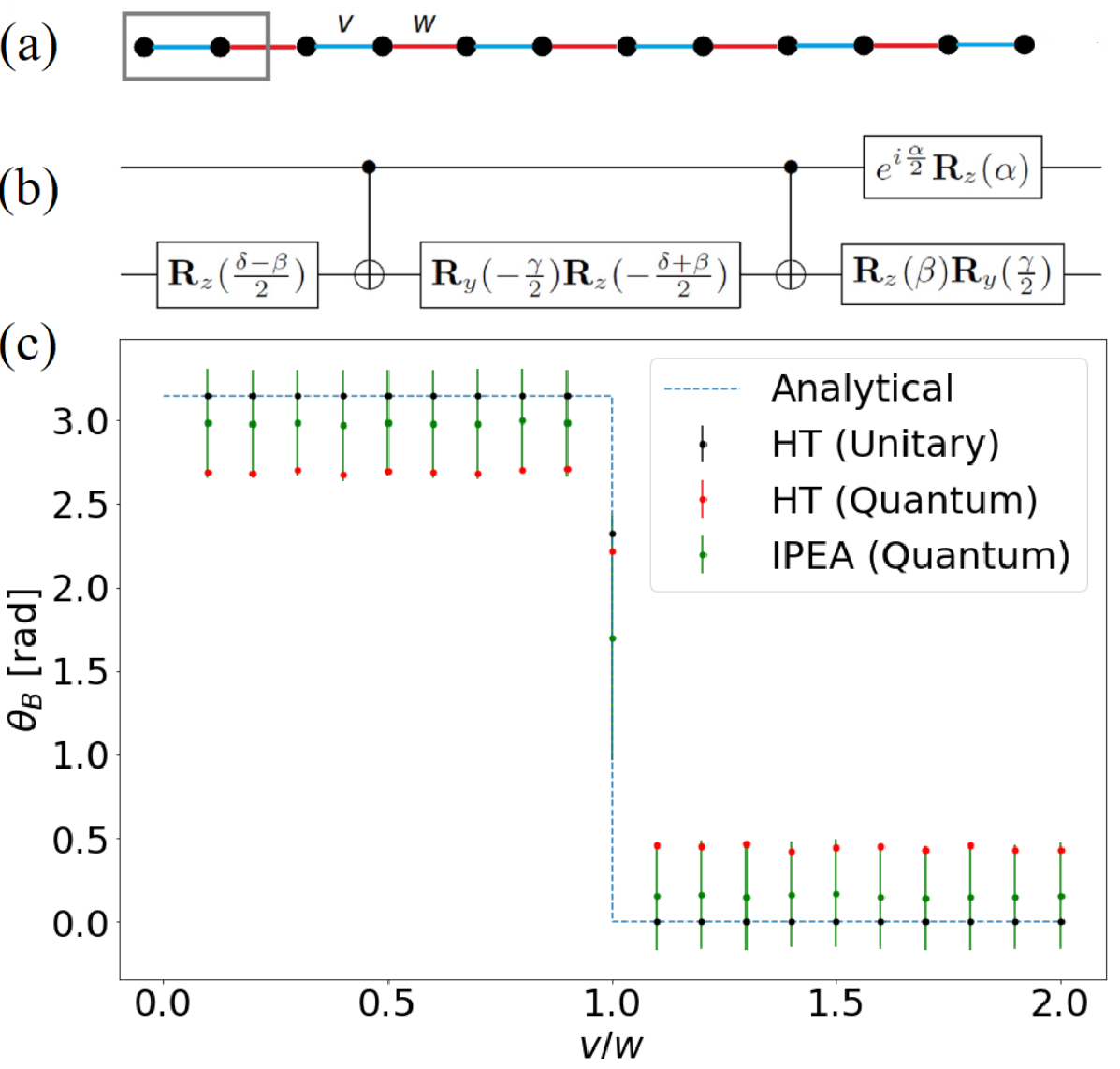}
\caption{(a) SSH tight-binding chain with intra- and inter-cell hopping parameters $v$ and $w$. Gray box delimits a unit cell. (b) Quantum circuit implementing controlled-$\delta \mathcal{U}$ gate defined in eq. (\ref{Uprod}). $\alpha$, $\beta$, $\gamma$ and $\delta$ are related to the parameter $\rho$ and the Hamiltonian $\mathcal{H}_{SSH}$ as described in the supp. mat. (c) Berry phase, as obtained from eq. (\ref{Z})  analytically (blue), \textit{in-silico} unitary simulation (black) and experimental quantum simulation (red) of the Hadamard-test (HT) circuit (Fig. 1(a)), and quantum simulation of the IPEA circuit (Fig. 1(b)) for $R = 4$ iterations and $N = 4$ time steps (green). Quantum simulations were carried out in the \textit{ibmq\_16\_melbourne} device from the IBM Q Experience. Further details about the quantum simulation can be found in the supp. mat.\label{fig2}}
\end{figure}

We now substantiate our proposal by describing the quantum circuit that implements $ {\cal U}_{\rm loop}$  in the context of topological classification of quantum phases of two different model Hamiltonians.  We first consider the SSH model, which describes a one-dimensional tight-binding model for spinless fermions with one orbital per site,  intra-cell hopping $v$ and inter-cell hopping $w$ (Fig. 2(a)). Using Bloch's theorem, the  Hamiltonian can be block-diagonalized in terms of $2 \times 2$ matrices:
%
%
\begin{equation}
\mathcal{H}_{SSH}(k) = (v + w \cos k) \; \sigma_x + (w \sin k) \; \sigma_y\equiv \vec{h}(k)\cdot\vec{\sigma},
\label{SSH}
\end{equation}

\noindent where $k$ is the wavenumber. For $v\neq w$ this model describes an insulator, with two energy bands separated by a gap of size $2 |v-w|$ at $k=\pm \pi$. The different topological nature of these two phases is characterized by a specific case of Berry phase, known as Zak phase \cite{zak1989}, which is obtained when the ground state of $\mathcal{H}_{SSH}(k)$ is looped in $k$-space across the 1\textsuperscript{st} Brillouin zone. The Berry phase can be used as a topological index:
\begin{equation}
\theta_B=
 -i \int_0^{2\pi} \langle \Psi_G(k)|\frac{\partial \Psi_G(k)}{\partial k}\rangle dk
=\begin{cases}
\pi &,\; v<w \\
0 &,\; v>w
\end{cases}
.
\label{Z}
\end{equation}
The number of in-gap edge modes is given by $2\theta_B/\pi$, so that only the $v<w$ phase has robust in-gap edge states and is said to be topological. This is a manifestation of the bulk-boundary correspondence \cite{hasankane2010,qizhang2011}. Hence, at $v = w$ there is a topological phase transition as a topological invariant changes value.   
  
Taking the SSH model to reciprocal space permits several simplifications. First, the wave function can be encoded in a single qubit. Second, the controlled unitary operations can be implemented by taking advantage of closed-form analytical expressions for the unitary evolution operator (see suppl. mat.). The  Berry phase for the SSH model as a function of $v/w$  is shown in  Fig. 2(c) as obtained  in four different ways: analytically (eq. (\ref{Z})), via an \textit{in-silico} simulation of the Hadamard test circuit shown in Fig. 1(a), and via the implementation of both circuits in Fig. 1 on the \textit{ibmq\_16\_melbourne} device
\footnote{\url{https://github.com/Qiskit/ibmq-device-information/tree/master/backends/melbourne/V1}}.
The controlled-$\delta \mathcal{U}$ gate was implemented using the circuit of Fig. 2(b) in both the unitary simulation and the actual quantum computations.
 
The results for the Hadamard-test circuit in quantum hardware (red markers) are close but not quite within (shot noise) error range from the analytical values for $N=4$ time steps. This is due to the limitations of current quantum hardware. We have verified that IPEA (Fig. 1b) gives results closer to the theory with $R=4$ iterations (see green markers in Fig. 2(c)). Naturally, for $R$ iterations the maximum precision that can be achieved is $2^{-R}$, while in the original method the precision is shot-noise bounded. The choice between the Hadamard-test circuit and IPEA therefore involves a trade-off between accuracy and precision.

The topological classification of non-interacting models can be efficiently done with classical computers. This is, however, {\em not} the case for interacting systems for which there are no analytical solutions and whose size is beyond the capacity of conventional computers. We now show that our algorithm can be used in this second class of non-trivial systems. To do so, we implement our proposal to address the topological classification of the ground state of the dimerized Heisenberg spin chain. The Hamiltonian for periodic boundary conditions (PBC) reads as:
\begin{equation} 
    \hat{H}_{\rm PBC} = \sum_{i = 0}^{N_s/2-1} \big( \frac{J_+}{4} \; \vec{\sigma}_{2i+1} \cdot \vec{\sigma}_{2i+2} +\frac{J_-}{4} \; \vec{\sigma}_{2i+2} \cdot \vec{\sigma}_{2i+3} \big), 
\end{equation}

\noindent where $J_{\pm}=J\pm \delta$, $J$ is the average spin coupling, $\delta$ is the dimerization parameter, $N_s$ is the number of $S = 1/2$ spins in the chain, $\vec{\sigma_i} = (\sigma_i^{x}, \sigma_i^{y}, \sigma_i^{z})^T$ is the Pauli vector for the i\textsuperscript{th} spin, and $\vec{\sigma}_{N_s + 1} = \vec{\sigma}_{1}$ due to the periodicity.
The Hamiltonian for open boundary conditions (OBC)  reads ${\cal H}_{\rm OBC}={\cal H}_{\rm PBC} -\frac{J_-}{4} \; \vec{\sigma}_N \cdot \vec{\sigma}_1$. This Hamiltonian has only been solved analytically for the case  $\delta=0$ \cite{bethe1931,bethe1997}, the well-known gapless spin liquid. For $\delta \neq 0$, reliable information is based on Density Matrix Renormalization Group \cite{lado2019} and exact diagonalizations \cite{fields1979}.  As for the SSH Hamiltonian, the OBC chain has in-gap edge excitations for $\delta<0$, but not for $\delta>0$. This, together with the fact that this model can be obtained from the SSH model when strong Hubbard repulsion is added \cite{Anderson59, Moriya60}, implies the two phases are topologically different.


The topological classification of the model can be done using a method proposed by Hatsugai \cite{Hatsugai2006}, which consists of introducing a twist phase $\rho$ in a \textit{single} local bond:
\begin{equation}
\vec{\sigma}_i \cdot \vec{\sigma}_j \rightarrow \frac{1}{2} (e^{-i\rho} \sigma_i^{+} \sigma_j^{-} + e^{i\rho} \sigma_i^{-} \sigma_j^{+}) + \sigma_i^z \sigma_j^z.
\end{equation}

\noindent The PBC ground state remains non-degenerate and gapped as $\rho$ is ramped between $0$ and $2\pi$ in the ring geometry. The Berry phase $\theta_B(j)$ that arises from this $\rho$-loop defines a local topological marker that reveals the dimer structure of the chain: it is $\pi$ at the stronger links and $0$ at the weaker ones (Fig. 3(a)). As $\delta$ changes sign and a strong bond becomes a weak one, the corresponding local Berry phase goes from $\pi$ to $0$, and vice-versa. Crucially, if a strong bond is removed from the PBC ring, the resulting  OBC chain is topologically non-trivial due to the presence of topologically-protected edge states. If instead a weak bond is removed, no edge states appear.

To implement the Berry phase estimation algorithm, several technical caveats that were absent in the case of the SSH model have to be dealt with. First, we need as many qubits as sites in the spin chain. Remarkably, the topological transition survives in small systems with as few as 4 spins, although finite size effects are present (see suppl. mat.). Second, the ground state $|\Psi_G (\rho = 0) \rangle$, which is no longer a product state in the computational basis, must be initialized before the start of the adiabatic loop. This is accomplished in two stages: obtaining $|\Psi_G (\rho = 0) \rangle$ as a linear combination of computational basis states via numerical diagonalization of the model, followed by the preparation of the state using the approach proposed  by Shende {\em et al.} \cite{shende2006}. The number of gates required for this initialization varies depending on the specific values of $J$, $\delta$ and $\rho$, taking values between 44 and 79 for a 4-spin ring. The number of CNOT gates corresponds to roughly two thirds of the total number of gates. Third, the Hamiltonian is the sum of non-commuting terms, so the implementation of the propagator requires a Trotter-Suzuki expansion \cite{Trotter,Suzuki}. Last, the decomposition of the controlled propagator in terms of basis gates cannot be achieved via the Z-Y-Z decomposition as before, since the input register involves more than one qubit. Instead, we make use of a scheme proposed in \cite{NielsenChuang} (see supp. mat.).

The results of the in-silico simulation of the Berry phase estimation algorithm (Hadamard test circuit shown in Fig. 1(a)) applied to the topological classification of the dimerized Heisenberg ring are shown in Fig. 3(b) (green markers), for a system with $N_s= 4$ spins and $J = 1$, as $\delta$ is ramped. These results differ only slightly from those obtained from the numerical simulation in classical hardware where the propagator of the full Hamiltonian is obtained via exact exponentiation (orange markers). This minor discrepancy is due to shot noise and Trotterization errors. Both sets of results deviate from the expected step-like pattern (blue dashed line) due to finite size effects (see supp. mat.). 

\begin{figure}[hbt]
\includegraphics[width=0.95\linewidth]{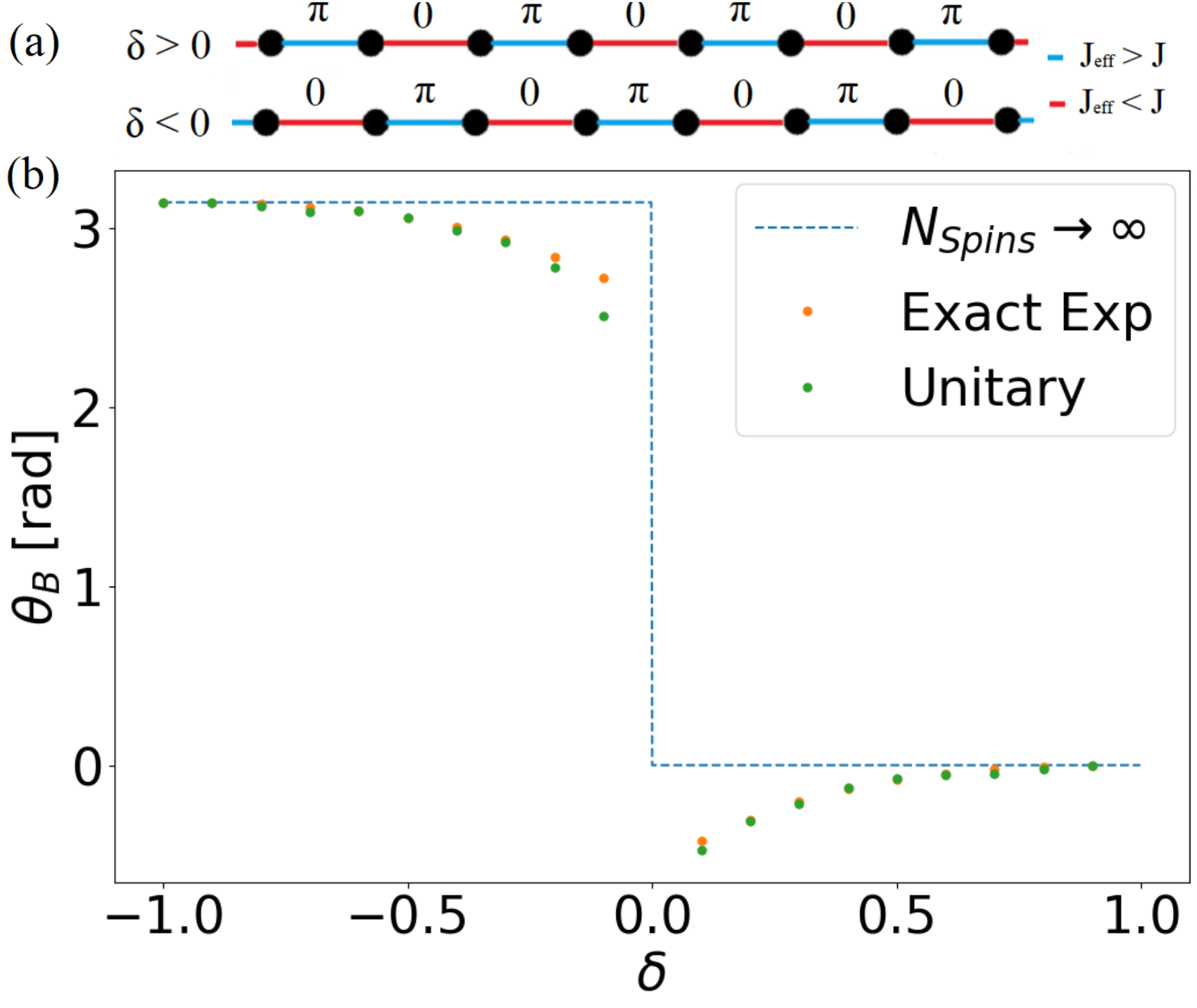}
\caption{(a) Local Berry phase $\theta_B$ obtained via Hatsugai twist \cite{Hatsugai2006} as a local topological marker that reveals the dimer structure of Heisenberg ring. Strong links ($J_{\textrm{eff}} \equiv J \pm \delta > J$) have $\theta_B = \pi$, while weak ones ($J_{\textrm{eff}} < J$) have $\theta_B = 0$. (b) Local Berry phase of a link with coupling $J_{\textrm{eff}} = J - \delta$ for a dimerized Heisenberg ring of 4 spins with $J = 1$ obtained via a noiseless unitary simulation of the Berry phase estimation circuit shown in Fig. 1(a). Quantum algorithm results (green markers) deviate slightly from those obtained via exact exponentiation of the full $16 \times 16$ Hamiltonian (orange markers) due to shot noise and Trotterization errors. Both simulations reveal a deviation from the expected step-like pattern (blue dashed line) due to finite size effects (see supp. mat.)
\label{fig3}}
\end{figure}

Finally, we discuss the perspective for implementation of the topological classification algorithm in quantum hardware. As a preliminary step, two sanity checks on the ground state initialization for a ring of 4 spins were carried out. First, its energy was estimated via IPEA \cite{Dobsicek} with $R=8$ in a noiseless simulation; good agreement with the exact diagonalization results was observed. Second, a parity conservation check \cite{McArdle2019} was conducted in both a noiseless simulation and the \textit{ibmq\_16\_melbourne} device. The results, shown in Fig. 6 of suppl. mat., show a clear discrepancy between the noiseless simulation and the actual quantum experiment, implying that the $\mathcal{U}_{\textrm{init}}$ subroutine alone is too deep for current quantum hardware even for just 4 spins.  

In principle, our Berry phase estimation quantum algorithm method can be used to implement the topological classification scheme proposed by Hatsugai \cite{Hatsugai2006} in higher-spin systems, in higher dimensions, and also for fermions in general \cite{le2019}. Hence, we hope that the proposed Berry phase estimation algorithm will be used as a tool to explore interacting topological phases when digital quantum computers outperform conventional computers in the simulation of quantum systems.

In summary, we have proposed a quantum algorithm to estimate the Berry phase acquired during the digitized quantum simulation of the ground state of an arbitrary Hamiltonian as it undergoes an adiabatic loop in a parameter space. Our approach combines phase estimation algorithms \cite{cleve98,Dobsicek} with the gate-based quantum simulation of cyclic adiabatic evolutions to estimate the Berry phase. We have discussed the use of this algorithm to classify topological phases of two types of Hamiltonians: the SSH model and the dimerized Heisenberg spin model in 1D. We have also successfully implemented the algorithm in IBM quantum hardware, evidencing the topological phase transition of the SSH chain. This work illustrates the potential of digital quantum computing to simulate topological quantum many-body systems.

\pagebreak

\onecolumngrid


\begin{center}
  \textbf{\large Supplemental Material for\\ ``Berry Phase Estimation in Gate-Based Adiabatic Quantum Simulation''}\\[.2cm]
\end{center}

\setcounter{equation}{0}
\setcounter{figure}{0}
\setcounter{table}{0}
\renewcommand{\theequation}{S\arabic{equation}}
\renewcommand{\thefigure}{S\arabic{figure}}

\section{I. \; Quantum Berry Phase Estimation Algorithm}

The Berry phase algorithm we propose resembles the simplest version of quantum phase estimation \cite{NielsenChuang}. As mentioned in the main text, however, there are two key differences: first, the input register in the Berry phase algorithm does not remain in the same state throughout the action of the $\mathcal{U}$ gate; second, a na\"{i}f implementation of the Berry phase algorithm yields the sum of the Berry and dynamical phases, so the latter must be cancelled out. 

In the quantum phase estimation algorithm, the input register is initialized in an eigenstate $\ket{\psi}$ of the Hamiltonian $\mathcal{H}$ \footnote{For the sake of completeness, it should be noted that Abrams and Lloyd \cite{abramslloyd1999} proposed initializing the input register in a state that overlaps with the desired eigenstate of the Hamiltonian. The register will then collapse onto the desired eigenstate with a probability given by the square of the degree of overlap. In this instance, the state of the input register also changes, but, contrary to our Berry phase estimation algorithm, only does so in the first iteration. From then on, the state will always remain the same.}. After applying a Hadamard gate to the ancilla qubit, the state of the ancilla + input register system is 

\begin{equation}
    \frac{\ket{0} + \ket{1}}{\sqrt{2}} \otimes \ket{\psi}.
\end{equation}

\noindent Being also an eigenstate of $\mathcal{U} = e^{-i \mathcal{H} t / \hbar}$, the right-hand side of this Kronecker product remains $\ket{\psi}$ throughout the entire circuit. Since the action of the propagator $\mathcal{U}$ is controlled by the ancilla qubit, the phase from which the eigenvalue can be obtained is only kicked back to the ancilla if it is in state $\ket{1}$: 

\begin{equation}
    \frac{\ket{0} + e^{-i E t / \hbar} \ket{1}}{\sqrt{2}} \otimes \ket{\psi},
\end{equation}

\noindent where $\mathcal{H} \ket{\psi} = E \ket{\psi}$. In the case of the Berry phase algorithm, although the input register is also initialized in an eigenstate (specifically, the ground state $\ket{\textbf{0}}$) of the starting Hamiltonian $\mathcal{H}(t = 0)$, at a later time the Hamiltonian $\mathcal{H}(t)$ is no longer the same as the starting one and, if the adiabatic condition is met, the input register will be in the (instantaneous) ground state of $\mathcal{H}(t)$, $\ket{\textbf{0'}}$, which is also different from $\ket{\textbf{0}}$. Hence, at an arbitrary time $t$ the wave function of the ancilla + input register system is

\begin{equation}
    \frac{\ket{0}}{\sqrt{2}} \otimes \ket{\textbf{0}} + \frac{e^{i\theta} \ket{1}}{\sqrt{2}} \otimes \ket{\textbf{0'}},
\end{equation}

\noindent where the phase $\theta$ includes both dynamic and geometric contributions. Because $\ket{\textbf{0}} \neq \ket{\textbf{0'}}$ for an arbitrary $t$, the ancilla and input register qubits are entangled, so the phase $\theta$ cannot be measured. It is only when the path in parameter space is closed --- and hence $t = T$, where $T$ is the period of the adiabatic cycle --- that the final state of the input register coincides with the initial:

\begin{equation}
    \frac{\ket{0}}{\sqrt{2}} \otimes \ket{\textbf{0}} + \frac{e^{i\theta} \ket{1}}{\sqrt{2}} \otimes \ket{\textbf{0}} = \Big( \frac{\ket{0} + e^{i\theta} \ket{1}}{\sqrt{2}} \Big) \otimes \ket{\textbf{0}}.
\end{equation}

\noindent The ancilla qubit is now disentangled from the input register, so we can proceed as in quantum phase estimation, obtaining the phase $\theta$ via a measurement in the Hadamard basis.

Let us now consider how to cancel the dynamical phase whilst keeping the Berry phase. Before discussing the quantum circuits that accomplish this, let us first analyze the effect of reversing the time arrow in the Berry and dynamical phases. If the adiabatic evolution is carried out backwards in time, the dynamical phase changes sign because the time step becomes negative --- $\theta_D' = \int_0^{-T} E(t) \; dt = \int_0^{T} E(t) \; (-dt) = -\theta_D$ ---, but the Berry phase remains invariant as it only depends on the rotation in parameter space. Hence, the dynamical and Berry phases are anti-symmetric and symmetric under time reversal, respectively.

Setting ${\cal U}_{\rm loop}=U_{\circlearrowleft}(0,T)$ in the circuit shown in Fig. 1(a) from the main text, after the action of the controlled-propagator $c \, \mathcal{U}_{\textrm{loop}}$, the state of the two qubits is

\begin{equation} \label{circuito_11}
    \Big( \frac{1}{\sqrt{2}}\ket{0} + \frac{e^{i(\theta_B + \theta_D)}}{\sqrt{2}}\ket{1} \Big) \otimes \ket{\textbf{0}},
\end{equation}

\noindent where $\ket{\textbf{0}}$ is the ground state of the given Hamiltonian (i.e. $\mathcal{U}_{init} \otimes_{i = 1}^n \ket{0} = \ket{\textbf{0}}$). The final Hadamard gate gives

\begin{equation} \label{circuito_12}
    \Big(\frac{1+e^{i(\theta_B + \theta_D)}}{2}\ket{0} + + \frac{1-e^{i(\theta_B + \theta_D)}}{2}\ket{1} \Big) \otimes \ket{\textbf{0}},
\end{equation}

\noindent and therefore the probability of measuring the ancilla qubit in state $\ket{0}$ is

\begin{equation} \label{circuito_13}
    P_0 = \left| \frac{1+e^{i(\theta_B + \theta_D)}}{2} \right|^2 = \cos^2\left(\frac{\theta_B + \theta_D}{2}\right).
\end{equation}

This circuit can be slightly modified in order to cancel the dynamical phase regardless of the choice of the number of time steps $N$ or the duration of the time step $\delta t$. This can be accomplished by setting ${\cal U}_{\rm loop}=U_{\circlearrowleft}(0,T) U_{\circlearrowleft}(T,0)$: the target qubit is evolved forward in time first, and then backward in time. The state of the two qubits after these two controlled-propagators is

\begin{equation} \label{circuito_21}
\begin{split}
    & \frac{1}{\sqrt{2}}\Bigg[ \Big( \ket{0} + \ket{1} \otimes (e^{i(\theta_D + \theta_B)})(e^{i(-\theta_D + \theta_B)}) \Big) \otimes \ket{\textbf{0}} \Bigg] = \\
    & = \frac{1}{\sqrt{2}}\Bigg[ \Big( \ket{0} + e^{i2\theta_B} \ket{1} \Big) \otimes \ket{\textbf{0}} \Bigg].
\end{split}
\end{equation}

\noindent After the Hadamard gate the state of the two qubits is

\begin{equation} \label{circuito_21}
    \Big( \frac{1+e^{i2\theta_B}}{2}\ket{0} + \frac{1-e^{i2\theta_B}}{2} \ket{1} \Big) \otimes \ket{\textbf{0}}
\end{equation}

\noindent and so the probability of a measurement of the ancilla qubit yielding 0 is $\cos^2\theta_B$.

Despite succeeding at cancelling the dynamical phase, this two-loop method has two main issues: first, since two loops are required instead of just one, the circuit is essentially twice as deep as the original proposal; second, the Berry phase is only defined within the range $[0, \pi)$, i.e. $0$ and $\pi$ are equivalent. The latter is a critical caveat, as several topological phase transitions involve a change of a topological parameter between $0$ and $\pi$.

Both issues are addressed by the performing a single adiabatic loop with ${\cal U}_{\rm loop}=U_{\circlearrowleft}(0,T/2) U_{\circlearrowleft}(T/2,0)$, i.e. the initial state is propagated forward in time during half of the loop and backward during the other half. If the energy spectrum is symmetric under reflection in the path corresponding to the adiabatic loop in parameter space, the dynamic phase cancels out. Hence, after the action of $c \; \mathcal{U}_{\textrm{loop}}$ the state of the two qubits is:

\begin{equation}
    \Big( \frac{1}{\sqrt{2}}\ket{0} + \frac{e^{i\theta_B}}{\sqrt{2}}\ket{1} \Big) \otimes \ket{\textbf{0}},
\end{equation}

\noindent in which case the probability of measuring the ancilla qubit in $\ket{0}$ is $\cos^2(\theta_B/2)$.

\section{II. \; Proof of Concept: Topological Phase Transition in Non-Interacting Model}

\subsection{i. \; Su-Schrieffer-Heeger (SSH) Model}

The bulk momentum-space SSH Hamiltonian is

\begin{equation}
    \mathcal{H}_{SSH}(k) = \left( \begin{matrix} 0 & v + w \: e^{-ik} \\ v + w \: e^{ik} & 0 \end{matrix} \right) \equiv \vec{d}(k) \cdot \vec{\sigma}, \quad \quad \vec{d}(k) = (v + w\cos k, w \sin k, 0)^T.
\end{equation}

\noindent The eigenstates of $\mathcal{H}_{SSH}(k)$ are given by $\ket{\psi_{\pm{}}} = \frac{1}{\sqrt{2}} (1, \; \pm{} e^{-i\phi(k)})^T$, where $\phi(k) = \textrm{Im} \{ \ln(v + w \; e^{-ik}) \}$, with eigenvalues $E_{\pm{}}(k) = \pm{} |v + w \: e^{-ik} | = \pm{} \sqrt{v^2 + w^2 + 2 v w \cos k}$. The corresponding Bloch vectors are $\uvec{h}_{\pm{}} = \pm{} (\cos \phi(k), \sin \phi(k), 0)$.

\begin{figure}[h]
\centering{\
\includegraphics[width=0.95\linewidth]{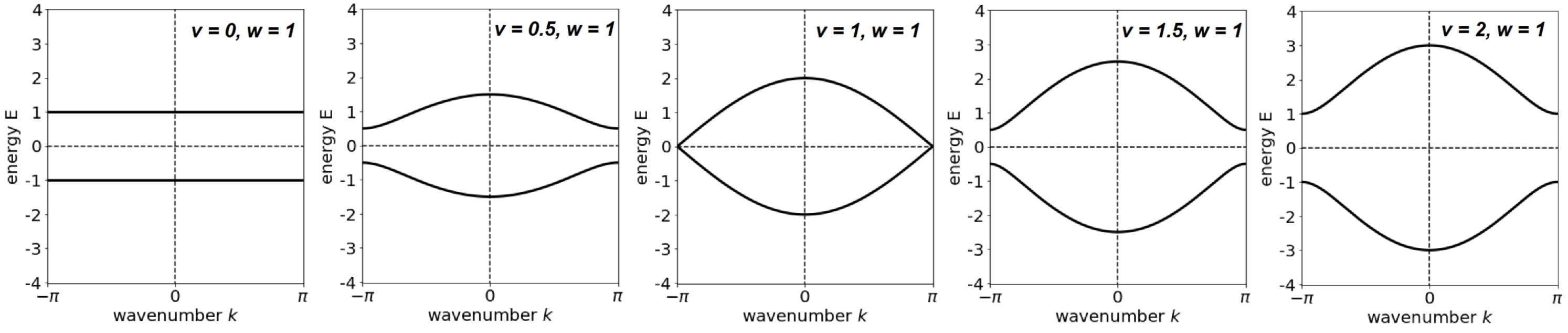}
\par}
\caption{Bulk band structure of SSH model for $w = 1$ and different values of $v$ in range $[0,2]$. Model describes a bulk insulator for all values of $v$ expect $v = 1$, at which the bulk gap closes at $k = \pm \pi$ and the system becomes a semi-metal. It is this closing of the bulk gap that allows the topological invariant $Z$ to change from 1 to 0
\label{loop}}
\end{figure}

An important property of $\mathcal{H}_{SSH}(k)$ is its chiral symmetry --- $\sigma_z \mathcal{H}_{SSH}(k) \sigma_z = - \mathcal{H}_{SSH}(k)$ ---, which imposes that the path traversed by the Bloch vectors of the eigenstates of $\mathcal{H}_{SSH}(k)$ as the wavenumber $k$ goes through the 1\textsuperscript{st} Brillouin zone is in the $d_x-d_y$ plane. This means that the winding number

\begin{equation}
    Z = \frac{1}{2\pi} \int_{-\pi}^{\pi} \Big( \uvec{h}_{-}(k) \times \frac{d}{dk} \uvec{h}_{-}(k) \Big)_z dk
\label{winding}
\end{equation}

\noindent is well-defined. Replacing $\uvec{h}_{-}$ in equation (\ref{winding}) gives:

\begin{equation}
     Z = -\frac{1}{2\pi} \int_{-\pi}^{\pi} dk \partial_k \phi(k)
    \label{winding_2}
\end{equation}

This winding number can be directly related to the Berry phase acquired by the ground state over this loop:

\begin{equation}
    \theta_B = i \oint d\vec{\textbf{d}} \bra{\psi_{-}} \vec{\nabla}_{\vec{\textbf{d}}} \ket{\psi_{-}} = i \int_{-\pi}^{\pi} w dk \bra{\psi_{-}} \frac{1}{w} \partial_k \ket{\psi_{-}} = -\frac{1}{2} \int_{-\pi}^{\pi} dk \partial_k \phi(k) = \pi \; Z
\end{equation}

\noindent where in the second equality we made use of the fact that the trajectory in $\vec{\textbf{d}}$-space is a circle of radius $w$ (Fig. \ref{loop}) and the factor of $1/2$ arises from the normalization of $\ket{\psi_{\pm{}}}$. Replacing $\phi(k) = \textrm{Im} \{ \ln(v + w \; e^{-ik}) \}$ in equation (\ref{winding_2}) gives:

\begin{equation}
    Z = - \frac{1}{2\pi} \int_{-\pi}^{\pi} dk \; \partial_k \Big( \textrm{Im} \big\{ \ln(v + w \; e^{-ik}) \big\} \Big) = - \textrm{Im} \bigg\{ \frac{1}{2\pi} \int_{-\pi}^{\pi} \frac{-iw \: e^{-ik} \: dk}{v + w \: e^{-ik}} \bigg\} = - \textrm{Im} \bigg\{ \frac{1}{2\pi} \oint_{\mathcal{C}} \frac{dz}{z} \bigg\}
\end{equation}

\noindent where in the last step the substitution $z = v + w \: e^{-ik}$ was used. $\mathcal{C}$ is the circular trajectory shown in Fig. \ref{loop} for different values of $v \in [0,2]$ and $w = 1$. If this circle encloses the origin, the integral will contain the corresponding pole, the residue of which gives rise to a winding number of 1. This occurs when $v < w$. If instead $v > w$, the function is analytic across the entire region enclosed by the circular path, therefore per Cauchy's theorem the winding number is $0$. This corresponds to the case $v > w$.

\begin{figure}[h]
\centering{\
\includegraphics[width=0.95\linewidth]{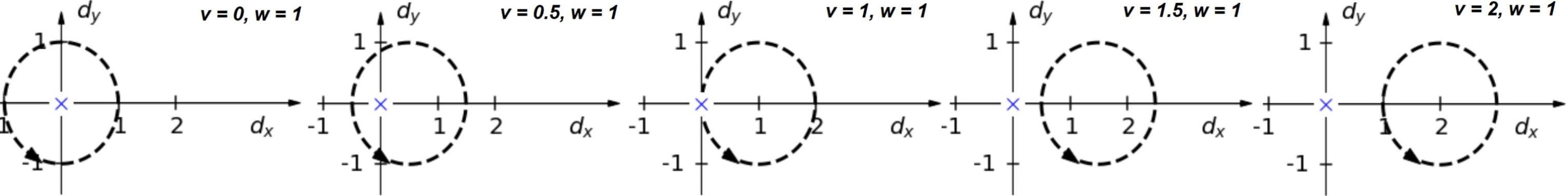}
\par}
\caption{Adiabatic loop in bulk momentum space of ground state of Su-Schrieffer-Heeger (SSH) Hamiltonian for different
values of staggered hopping amplitudes $v$ and $w$. From left to right, $v$ increases from $0$ to $2$ while $w$ remains fixed at $1$. When $v < w$, loop encloses degeneracy point $\vec{d} = 0$ (marked as a blue cross) and the ground state acquires a Berry phase of $\pi$ or, equivalently, a winding number $Z = 1$. This corresponds to the topological phase. For $v > w$, the Berry phase acquired over the loop is $0$, corresponding to a winding number $Z = 0$. This is the normal phase. Winding number is ill-defined at $v = w$
\label{loop}}
\end{figure}

The final note about the SSH model concerns the connection between its topological properties and the winding number. Indeed, the winding number is of great physical significance in the SSH model, as it corresponds to the net number of topologically protected edge states \cite{delplace2011}. This is an example of one of the most relevant features of topological insulators: the \textit{bulk-boundary correspondence} \cite{hasankane2010, qizhang2011}. It posits that the multiplicities of edge modes are equal to topological invariants of the bulk bands. A mathematically rigorous proof of the bulk-boundary correspondence for topological insulators is an open problem, although it has been corroborated by multiple experimental and numerical results in a wide range of contexts.

\subsection{ii. \; Implementation of Quantum Circuit}

\subsubsection{Ground State Initialization}

The ground state of the bulk momentum-space SSH Hamitonian $\mathcal{H}_{SSH}(k)$ is $\ket{\textbf{0}} \equiv \ket{\psi_{-}} = \frac{1}{\sqrt{2}} \; (1, \: - e^{i \phi(k)})^T$. The target qubit must be initialized in $\ket{\textbf{0}}$ before the start of the adiabatic evolution. This corresponds to the $\mathcal{U}_{init}$ operation shown on Figs. 1 and 2 from the main text. 

In practice, $\mathcal{U}_{init}$ was implemented via the IBM Quantum Information Science Kit (QISKit) \texttt{qiskit.extensions.initializer.initialize} function \footnote{\url{https://qiskit.org/documentation/autodoc/qiskit.extensions.initializer.html?highlight=initialize#qiskit.extensions.initializer.initialize}}, which follows a proposal by Shende, Bullock and Markov \cite{shende2006}. Starting from the desired state $\ket{\psi}$, this QISKit built-in function finds the circuit that maps it to the fiducial state $\ket{0}^{\otimes^n}$, where $n$ is the number of qubits required to encode $\ket{\psi}$. The initialization sub-circuit $\mathcal{U}_{init}$ is then the inverse of this circuit.

\subsubsection{Decomposition of Controlled-Propagator in Terms of Basis Gates}

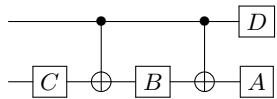
\begin{figure}
\centering{\
\Qcircuit @C=1em @R=1.2em {
  \lstick{} & \qw & \ctrl{1} & \qw & \ctrl{1} & \gate{D} \\
  \lstick{} & \gate{C} & \targ & \gate{B} & \targ & \gate{A} 
} \par}
\caption{Quantum circuit implementing controlled-$\mathcal{U}$ gate for $\mathcal{U} = e^{i\alpha} A X B X C, \; ABC = \mathbb{1}$, where $A = \textbf{R}_z(\beta)\textbf{R}_y(\gamma/2)$, $B = \textbf{R}_y(-\gamma/2) \textbf{R}_z(-(\delta + \beta)/2)$, $C = \textbf{R}_z((\delta - \beta)/2)$, and $D = e^{i\alpha/2} \textbf{R}_z(\alpha)$.}
\label{circuit_6}
\end{figure}

During the adiabatic loop in $k$-space, the wavenumber $k$ varies from $0$ to $2\pi$. Since we make use of gate-based quantum computers, the adiabatic evolution must be discretized in $N$ steps, each lasting $dt$. The parameter $k$ increases in steps of $2\pi/N$, being updated between consecutive time steps. The effective gate $\mathcal{U}$ is thus given as $\mathcal{U} = \prod_{j = 1}^N \mathcal{U}_j$, where $\mathcal{U}_j$ is given by

\begin{equation} \label{prop_form}
    \begin{split}
    &\mathcal{U}_j = \exp\bigg(-\frac{i}{\hbar}\mathcal{H}_{SSH}\Big(\frac{2\pi j}{N}\Big) \; \delta t\bigg) = \cos\bigg(\Big|\vec{h}\Big(\frac{2\pi j}{N}\Big)\Big| \; \delta t\bigg) \ \mathbb{1} - i \sin\bigg(\Big|\vec{h}\Big(\frac{2\pi j}{N}\Big)\Big| \; \delta t\bigg) \ \uvec{h}\Big(\frac{2\pi j}{N}\Big) \cdot \vec{\sigma} \\
    & = \left( \begin{matrix} \cos(|\vec{h}(\frac{2\pi j}{N})| \; \delta t) & -i \sin(|\vec{h}(\frac{2\pi j}{N})| \; \delta t) (v + w \; e^{-i\; 2\pi j/N}) \\ -i \sin(|\vec{h}(\frac{2\pi j}{N})| \; \delta t) (v + w \; e^{i \; 2\pi j/N}) & \cos(|\vec{h}(\frac{2\pi j}{N})| \; \delta t) \end{matrix} \right) \\
    & \equiv \left( \begin{matrix} a_j(2\pi j/N) & b_j(2\pi j/N) \\ -b_j^{*}(2\pi j/N) & a_j^{*}(2\pi j/N) \end{matrix} \right),
    \end{split}
\end{equation}

\noindent where $|\vec{h}(k)| = \sqrt{v^2 + w^2 + 2 v w \cos k}$ and $k = 2\pi j/N$ at step $j$. To implement this propagator in a real device, each of these infinitesimal elements must be decomposed into a sequence of elementary quantum gates. This can be accomplished through the Z-Y-Z decomposition \cite{NielsenChuang}:

\begin{equation} \label{zyz}
\begin{split}
    & \mathcal{U}_j = e^{i\alpha_j} \textbf{R}_z(\beta_j) \textbf{R}_y(\gamma_j) \textbf{R}_z(\delta_j) = \\
    & = \left( \begin{matrix} e^{i(\alpha_j - \beta_j/2 - \delta_j/2)} \cos\frac{\gamma_j}{2} & -e^{i(\alpha_j - \beta_j/2 + \delta_j/2)} \sin\frac{\gamma_j}{2} \\ e^{i(\alpha_j + \beta_j/2 - \delta_j/2)} \sin\frac{\gamma_j}{2} & e^{i(\alpha_j + \beta_j/2 + \delta_j/2)} \cos\frac{\gamma_j}{2} \end{matrix} \right),
\end{split}
\end{equation}

\noindent where $\textbf{R}_z(\beta)$ is the rotation matrix about the z-axis by angle $\beta$. Comparing equations (\ref{prop_form}) and (\ref{zyz}) gives

\begin{equation}
\begin{split}
    & \alpha_j = 0, \quad \beta_j = \pi - \textrm{arg}(b_j) - \textrm{arg}(a_j), \quad \\
    & \gamma_j = 2 \arctan\Big(\frac{|b_j|}{|a_j|}\Big), \quad \delta_j = -\pi -\textrm{arg}(a_j) + \textrm{arg}(b_j),
\end{split}
\end{equation}

\noindent where $a = |a| e^{i\textrm{arg}(a)}$ The advantage of the Z-Y-Z decomposition is the ease with which a single-qubit gate can be converted into a controlled one \cite{NielsenChuang}, as shown in Fig. \ref{circuit_6}.

\subsubsection{Device Structure and Qubit Properties}

The experiments were carried out in the \textit{ibmq\_16\_melbourne} device from the IBM Q Experience. The connectivity on the device is provided by 22 coplanar waveguide "bus" resonators, each of which connects two qubits. The coupling scheme, as well as further device specifications, can be found on \footnote{\url{https://github.com/Qiskit/ibmq-device-information/tree/master/backends/melbourne/V1}}.

The quantum circuit for the SSH model only makes use of two qubits. Hence, of the 14 made available in the \textit{ibmq\_16\_melbourne} device, we chose the coupled pair with the best combination of high coherence times $T_1$ and $T_2$, low cX gate errors and low readout errors for the ancilla qubit. According to the measurements from the previous calibration conducted by IBM, the cX error rate was of 3.7\%, the coherence times of the ancilla qubit were $T_1 = 64 \; \mu s$ and $T_2 = 97 \; \mu s$ and for the target qubit $T_1 = 67 \; \mu s$ and $T_2 = 124 \; \mu s$, and the readout error of the ancilla qubit 7.3\%.

\subsection{iii. \; Statistical Analysis}

In the Hadamard test circuit shown in Fig. 1(a) from the main text, the number of trials in each run of the experiment (i.e. for each value of the hopping parameter $v$ in the SSH model) was chosen to be 8192 --- which corresponds to the maximum allowed by the IBM Q Experience hardware for a single experiment --- to minimize shot noise. The error bars associated with this shot noise were estimated as $\sqrt{P(0) (1 - P(0))}/\sqrt{N}$, as expected for a binomial distribution. 

\begin{table}[htb]

    \centering
    \caption{Results of iterative phase estimation of Berry phase of ground state of SSH Hamiltonian for $v = 2, w = 1$. Berry phase was measured to four significant bits, the first (1\textsuperscript{st}) being the most significant. Most likely bitwise representation, $2\pi0.0000 = 0$, is in agreement with the expected value for the normal phase ($v > w$).
    }

    \begin{tabular}{ | p{2cm} | p{2cm} | p{2cm} | p{2cm} |}
    \hline
    \textbf{Bit} & \textbf{$\ket{0}$ Shots} & \textbf{$\ket{1}$ Shots} & \textbf{Expected} \\ \hline
    \textbf{1\textsuperscript{st}} & 7731 & 461 & 0 \\ \hline
    \textbf{2\textsuperscript{nd}} & 7614 & 578 & 0 \\ \hline
    \textbf{3\textsuperscript{rd}} & 7442 & 750 & 0 \\ \hline
    \textbf{4\textsuperscript{th}} & 6953 & 1239 & 0 \\
    \hline
    \end{tabular}
\end{table}

\tikzstyle{level 1}=[level distance=2.5cm, sibling distance=6cm]
\tikzstyle{level 2}=[level distance=2.5cm, sibling distance=3cm]
\tikzstyle{level 3}=[level distance=2.5cm, sibling distance=1.5cm]
\tikzstyle{level 4}=[level distance=2.5cm, sibling distance=0.5cm]
\tikzstyle{bag} = [circle,draw]
\tikzstyle{square} = [rectangle,draw]

\begin{figure}[h]
\centering
\begin{tikzpicture}[grow=right, sloped]
\node {}
    child {
        node[bag] {1}        
            child {
                node[bag] {11}
                    child {
                        node[bag] {111} 
                        child{
                            node[label=right:
                                {1111}] {}
                            edge from parent
                            node[below] {1/2}
                        }
                        child{
                            node[label=right:
                                {0111}] {}
                            edge from parent
                            node[above] {1/2}
                        }
                        edge from parent
                        node[below]  {1/2}
                    }
                    child {
                        node[bag] {011} 
                        child{
                            node[label=right:
                                {0011}] {}
                            edge from parent
                            node[below] {1/2}
                        }
                        child{
                            node[label=right:
                                {1011}] {}
                            edge from parent
                            node[above] {1/2}
                        }
                        edge from parent
                        node[above] {1/2}
                    }
                edge from parent
                node[below]  {1/2}
            }
            child {
                node[bag] {01}
                    child {
                        node[bag] {101} 
                        child{
                            node[label=right:
                                {1101}] {}
                            edge from parent
                            node[below] {1/2}
                        }
                        child{
                            node[label=right:
                                {0101}] {}
                            edge from parent
                            node[above] {1/2}
                        }
                        edge from parent
                        node[below]  {1/2}
                    }
                    child {
                        node[bag] {001} 
                        child{
                            node[label=right:
                                {1001}] {}
                            edge from parent
                            node[below] {1/2}
                        }
                        child{
                            node[label=right:
                                {0001}] {}
                            edge from parent
                            node[above] {1/2}
                        }
                        edge from parent
                        node[above] {1/2}
                    }
                edge from parent
                node[above]  {1/2}
            }
            edge from parent
            node[below] {1239/8192}
    }
    child {
        node[bag] {\textbf{0}}        
            child {
                node[bag] {10}
                    child {
                        node[bag] {110} 
                        child{
                            node[label=right:
                                {1110}] {}
                            edge from parent
                            node[below] {1/2}
                        }
                        child{
                            node[label=right:
                                {0110}] {}
                            edge from parent
                            node[above] {1/2}
                        }
                        edge from parent
                        node[below]  {1/2}
                    }
                    child {
                        node[bag] {010} 
                        child{
                            node[label=right:
                                {1010}] {}
                            edge from parent
                            node[below] {1/2}
                        }
                        child{
                            node[label=right:
                                {0010}] {}
                            edge from parent
                            node[above] {1/2}
                        }
                        edge from parent
                        node[above] {1/2}
                    }
                edge from parent
                node[below]  {750/8192}
            }
            child {
                node[bag] {\textbf{00}}
                    child {
                        node[bag]{100} 
                        child {
                            node[label=right:
                                {1100}] {}
                            edge from parent
                            node[below] {1/2}
                        }
                        child {
                            node[label=right:
                                {0100}] {}
                            edge from parent
                            node[above] {1/2}
                        }
                        edge from parent
                        node[below]  {578/8192}
                    }
                    child {
                        node[bag]{\textbf{000}} 
                        child{
                            node[label=right:
                                {1000}] {}
                            edge from parent
                            node[below] {461/8192}
                        }
                        child{
                            node[label=right:
                                {\textbf{0000}}] {}
                            edge from parent
                            node[above] {7731/8192}
                        }
                        edge from parent
                        node[above] {7614/8192}
                    }
                edge from parent
                node[above]  {7442/8192}
            }
            edge from parent
            node[above] {6953/8192}
    };
    
\end{tikzpicture}
\caption{Tree diagram of probability distribution for iterative phase estimation of Berry phase of ground state of SSH Hamiltonian for $v = 2, w = 1$, which corresponds to the normal phase. The measurement outcomes of the four iterations of the algorithm are shown in Table I. Significance of bits grows from left to right, that is, the first branch corresponds to the measurement of the 4\textsuperscript{th} and least significant bit. Binary sequences highlighted in bold were selected in each iteration. Since the probability of the discarded bit in the previous step is nonzero, unobserved sequences cannot be ignored in the statistical analysis. Unobserved sequences that differ by one bit are assumed to be equally likely.}
\label{tree}
\end{figure}
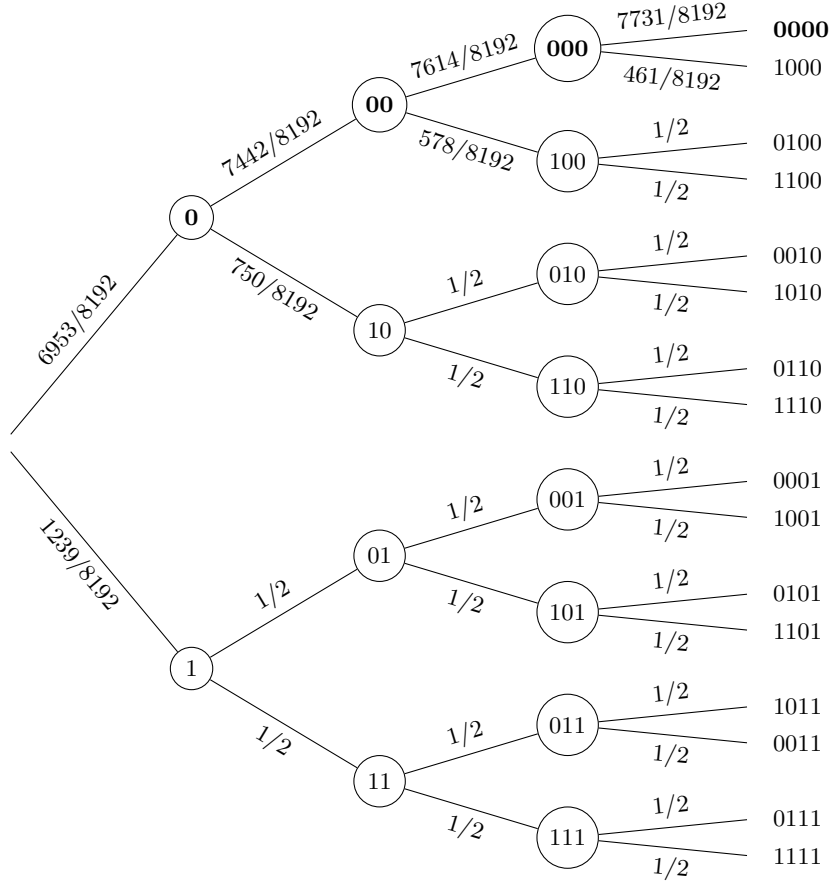

Regarding the iterative phase estimation, to determine the error associated with this measurement we followed the method proposed in \cite{Cruz_thesis}, which makes use of circular data statistics. The basic idea behind circular data statistics is that the average of the measured phases can be performed by summing complex numbers whose magnitude is given by the probability and whose complex phase is the phase itself, i.e.:

\begin{equation}
    \bar{R} = \sum_{k=0}^{2^n-1} P(\phi_k) e^{i2\pi\phi_k} = |\bar{R}| e^{i2\pi\hat{\phi}}.
\end{equation}

\noindent $\hat{\phi}$ is the expected value of the phase. $|\bar{R}|$ is a measure of variance: if $|\bar{R}|$ is close to 1, then all vectors have similar phase, leading to constructive interference; if instead $|\bar{R}|$ is close to 0, the measured phases are disperse, so the variance is large. The circular variance can therefore be defined as $V = 1 - |\bar{R}|$. The circular standard phase deviation $v$ is:

\begin{equation}
    v = \frac{\sqrt{-2\ln|\bar{R}|}}{2\pi}
\end{equation}

\noindent and can be taken as the error in the measurement.

There is, however, one caveat to the implementation of this statistical analysis for IPE: once a bit is measured, binary sequences that do not include that bit are discarded. For example, if the 4\textsuperscript{th} most significant bit is found to be 0, then all measured binary sequences will be of the form $xxx0$, where $x \in \{0,1\}$, which means that sequences such as $1111$ are not accounted for. To have a well-defined probability distribution (i.e. one that gives a value for each and every binary sequence) we assume that unobserved sequences that differ by only one bit have equal probability. This is schematically represented in Fig. \ref{tree} for the SSH model parameter values $(v,w) = (2,1)$, which corresponds to the normal phase of the SSH model. The bit sequence selected in each trial is highlighted in bold. The most likely bitwise representation of the Berry phase in this case is thus $2\pi0.0000 = 0$ with probability $\frac{6953}{8192} \; \frac{7442}{8192} \; \frac{7614}{8192} \; \frac{7731}{8192} \approx 0.68$. For the sake of clarity, the probability of the bitwise representation of the Berry phase being, for example, $2\pi0.1110 = \frac{7\pi}{4}$ is $\frac{6953}{8192} \; \frac{750}{8192} \; \frac{1}{2} \; \frac{1}{2} \approx 0.02$.

It is worth highlighting the decrease in probability of measuring the expected bit $0$ as the significance of the bit decreases (Table I). This is due to the fact that less significant bits require deeper circuits to be measured via IPE, which leads to a greater accumulation of errors in noisy quantum devices.

\section{III. \; Topological Characterization of Interacting Spin Model}

\subsection{i. \; Implementation of Quantum Circuit}

\subsubsection{Ground State Initialization}

The ground state of the dimerized Heisenberg model in a one-dimensional ring of 4 spins is computed via exact diagonalization using the QuSpin library \footnote{\url{http://weinbe58.github.io/QuSpin/}}. The 4-qubit register is then initialized in this state through the \texttt{qiskit.extensions.initializer.initialize} function already used for the SSH ground state initialization. The $\mathcal{U}_{init}$ subcircuit is, however, much deeper in this instance. The total number of gates varies depending on the specific values of $J$, $\delta$ and $\theta$, taking values between 44 and 79, two thirds of which are CNOTs.

\begin{figure}[h]
\centering{\
\includegraphics[width=14cm]{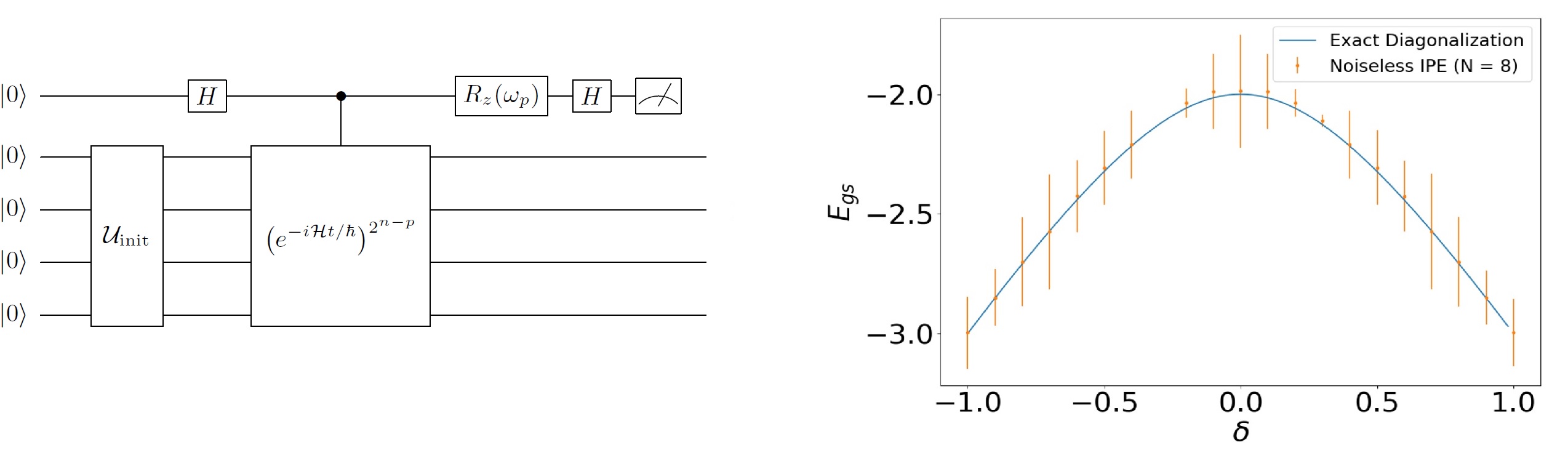}
\par}
\caption{Energy check of ground state of dimerized Heisenberg ring of 4 spins obtained from \texttt{qiskit.extensions.initializer.initialize} function. (Left) Circuit scheme of p\textsuperscript{th} iteration of iterative phase estimation algorithm (IPE) \cite{Dobsicek} to measure the ground state energy $E_{\textrm{gs}}$. $\mathcal{H}$ is the dimerized Heisenberg Hamiltonian for a ring of 4 spins. The $R_z(\omega_p)$ gate, where $\omega_p = -2\pi0.0...\phi_{p+1}...\phi_n$, that is applied after controlled-propagator serves to remove the contribution to the phase from the previously measured bits. The measured phase, $2\pi 0.\phi_1 \phi_2 ... \phi_n$, corresponds to $-E_{\textrm{gs}} t / \hbar$. The time variable $t$ is varied and the ground state energy $E_{\textrm{gs}}$ is obtained as the slope of the linear regression (with $\hbar = 1$, for convenience). (Right) Ground state energy measured via IPE on a noiseless \textit{in silico} simulator. Energy was measured to eight binary digits of precision (i.e. number of iterations was $n = 8$). Simulation results are in agreement with ground state energy obtained via exact diagonalization. 
\label{energy}}
\end{figure}

\begin{figure}[h]
\centering{\
\includegraphics[width=12cm]{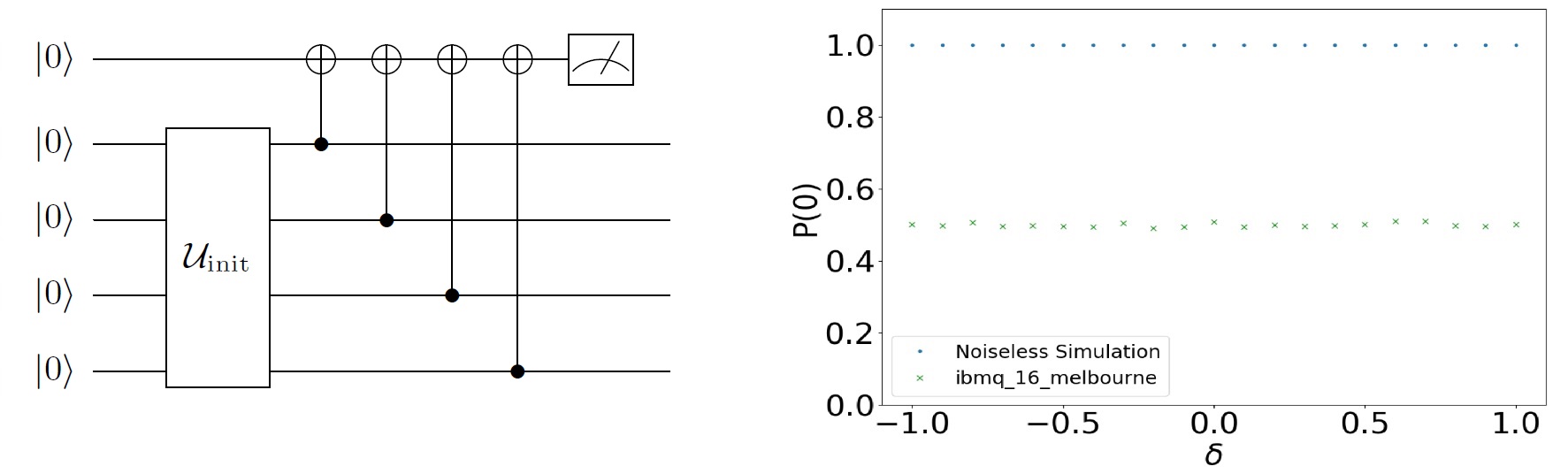}
\par}
\caption{Parity check of ground state of dimerized Heisenberg ring of 4 spins obtained from \texttt{qiskit.extensions.initializer.initialize} function. (Left) Circuit that measures parity of state initialized by $\mathcal{U}_{\textrm{init}}$ subroutine. The ancilla qubit at the top should be measured in the $\ket{\frac{1}{2} (1 - (-1)^4)} = \ket{0}$, since the system comprises 4 fermions (spin-$\frac{1}{2}$ particles) (Right) Outcome of implementation of parity check circuit in noiseless \textit{in silico} simulator and in \textit{ibmq\_16\_melbourne} device. Results of noiseless simulation are as expected, confirming that the ground state is initialized accurately. Implementation in real device, on the other hand, shows that parity check is far from verified, meaning that the ground state initialization alone yields too deep a circuit for state-of-the-art quantum hardware.
\label{parity}}
\end{figure}

Two sanity checks were performed to confirm the initialization of the ground state. First, the energy of the ground state was measured via iterative phase estimation in the noiseless simulator (Fig. \ref{energy}), being in agreement with the exact value. Second, the parity \cite{McArdle2019} was confirmed to be consistent with that of the exact ground state for a noiseless simulation, but not for an experiment in the \textit{ibmq\_16\_melbourne} device (Fig. \ref{parity}), so the ground state initialization alone already gives rise to too deep a circuit for current quantum hardware even for a ring with just 4 spins.

\subsubsection{Time Discretization and Decomposition of Controlled Propagator in Terms of Basis Gates}

Contrary to the SSH Hamiltonian, the Heisenberg Hamiltonian involves non-commuting terms. According to the Baker-Campbell-Hausdorff formula \cite{NielsenChuang}, $e^{(A+B)\delta t} = e^{A \delta t} e^{B \delta t} e^{-\frac{1}{2} [A,B] \delta t^2} + \mathcal{O} (\delta t^3)$. Hence, for non-commuting operators, the exponential of the sum is not equal to the product of the exponentials. To obtain the exponential of Hamiltonians involving non-commuting terms, assuming the exponential of each individual term can be computed, one must carry out a Trotter-Suzuki expansion \cite{Trotter, Suzuki}.

\begin{figure}[h]
\centering{\
\includegraphics[width=18cm]{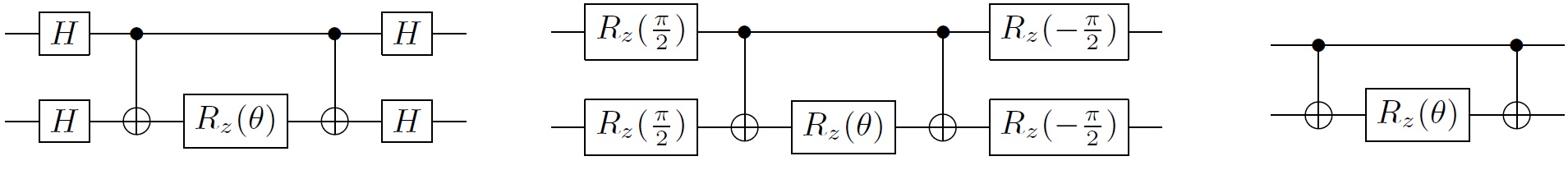}
\par}
\caption{Decomposition of $e^{-i\frac{\theta}{2}\sigma_i^x \sigma_{i+1}^x}$ (left), $e^{-i\frac{\theta}{2}\sigma_i^y \sigma_{i+1}^y}$ (center), and $e^{-i\frac{\theta}{2}\sigma_i^z \sigma_{i+1}^z}$ (right) exchange interactions in terms of basis gates. These are all the non-commuting interactions required to implement the dimerized Heisenberg model. 
\label{basis_gates_1}}
\end{figure}

\begin{figure}[h]
\centering{\
\includegraphics[width=16cm]{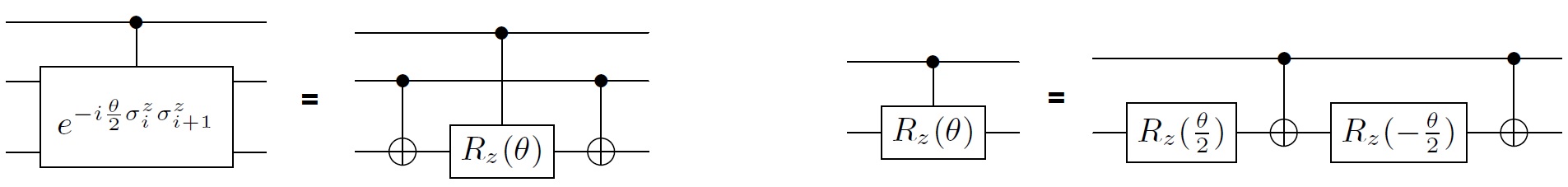}
\par}
\caption{Outline of implementation of controlled-propagators. (Left) Decomposition of controlled-$e^{-i\frac{\theta}{2}\sigma_i^z \sigma_{i+1}^z}$ in terms of basis gates and $cR_z$. (Right) Decomposition of $cR_z$ in terms of basis gates only
\label{basis_gates_2}}
\end{figure}

The Heisenberg Hamiltonian involves only three different terms: $\sigma_i^x \sigma_{i+1}^x$, $\sigma_i^y \sigma_{i+1}^y$ and $\sigma_i^z \sigma_{i+1}^z$. Each of these terms can be decomposed in terms of basis gates as shown in Fig. \ref{basis_gates_1}. To implement the controlled versions of these gates, the $R_z$ gate between the two CNOTs must be replaced by a $cR_z$, as illustrated on the left-hand side of Fig. \ref{basis_gates_2}. The right-hand side of Fig. \ref{basis_gates_2} shows the decomposition of the $cR_z$ gates in terms of basis gates.

However, the Hatsugai twist complicates the decomposition of the exchange interactions in the xy-plane. Indeed, $\frac{1}{2} (e^{-i\theta} \sigma_i^{+} \sigma_{i+1}^{-} + e^{i\theta} \sigma_i^{-} \sigma_{i+1}^{+}) = \cos \theta \; (\sigma_i^{x} \sigma_{i+1}^{x} + \sigma_i^{y} \sigma_{i+1}^{y}) + \sin \theta \; (\sigma_i^{y} \sigma_{i+1}^{x} - \sigma_i^{x} \sigma_{i+1}^{y})$, so the number of non-commuting terms becomes 5 instead of 3 for $\theta = 0$ case, which makes the Trotter-Suzuki expansion more convoluted. The number of time steps $N$ and the number of Trotter steps $N_{\textrm{Trotter}}$ within each time step were chosen to find complete agreement with the results obtained in classical hardware by obtaining the exponential of the full $16 \times 16$ Hamiltonian explicitly. Specifically, for the results shown in Fig. 3(b) from the main text, $N = 100$ and $N_{\textrm{Trotter}} = 10$.

\subsubsection{Finite Size Effects}

\begin{figure}[h]
\centering{\
\includegraphics[width=10cm]{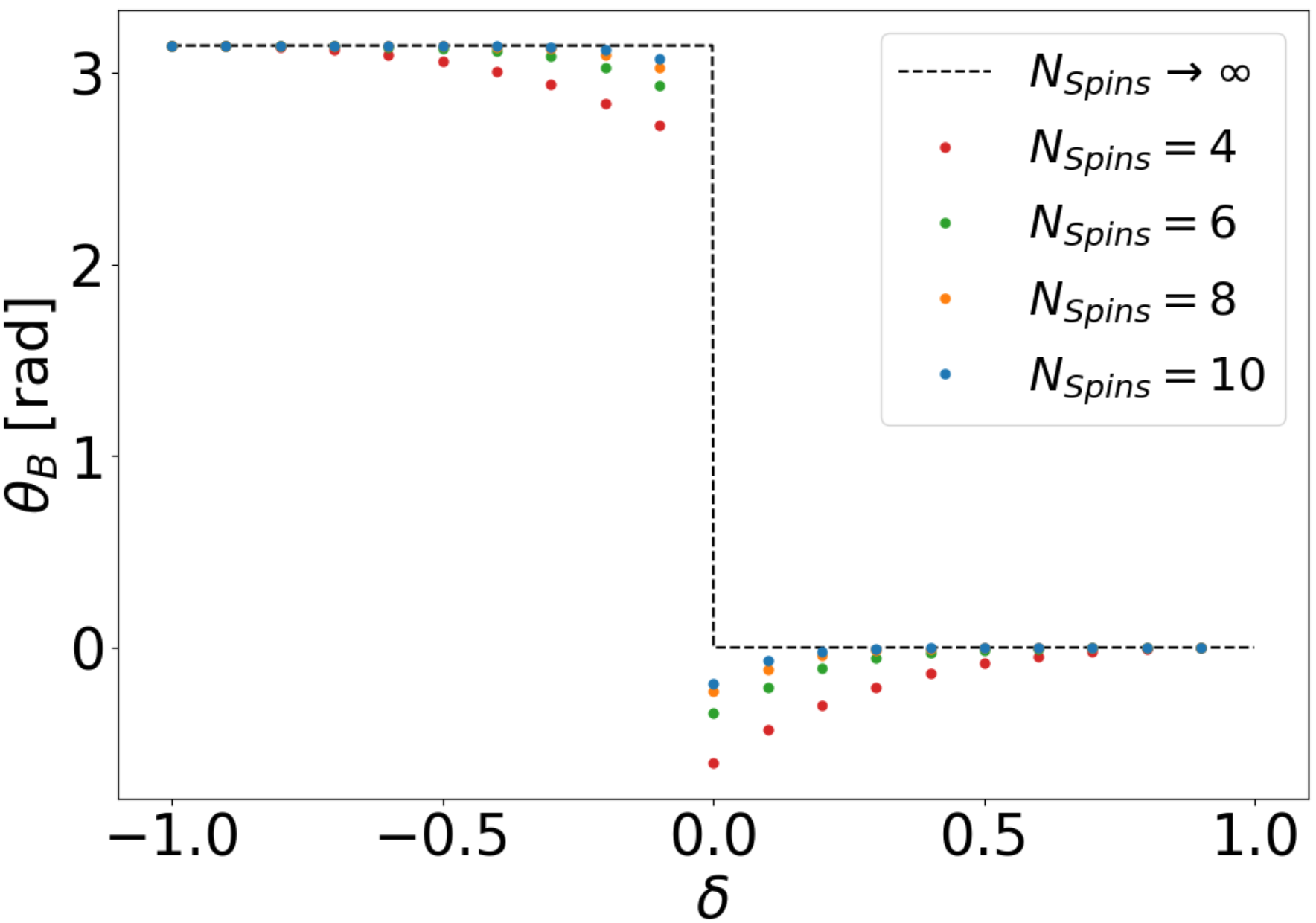}
\par}
\caption{Finite size effects in the measurement of the local Berry phase of a link with coupling $J_{\textrm{eff}} = J - \delta$ in a dimerized Heisenberg ring following Hatsugai's proposal \cite{Hatsugai2006}. Results were obtained in classical hardware by perfoming exponential of full $16 \times 16$ Hamiltonian explicitly. As the size of the ring is increased, the local Berry phase deviates less from $\pi$ for $\delta < 0$ and from $0$ for $\delta > 0$.
\label{finite_size}}
\end{figure}

As shown in Fig. 3(b) from the main text, the results of both the unitary simulation of the quantum circuits and the classical simulation via exact exponentiation of the full $16 \times 16$ Hamiltonian deviate from the expected step-like pattern. This is due to finite size effects, as shown below in Fig. \ref{finite_size}. Indeed, as the size of the ring is increased, the local Berry phase deviates less from $\pi$ for $\delta < 0$ and from $0$ for $\delta > 0$.


\pagebreak

{\it Acknowledgements}. We acknowledge Pedro Cruz and Jos\'{e} Luis Lado  for fruitful discussions. B.M. and J.F.R. acknowledge the FCT \textit{Functionalized Graphene for Quantum Technologies} project (PTDC/FIS-NAN/4662/2014). G.C. acknowledges the FCT PhD scholarship no. SFRH/BD/138806/2018. All authors acknowledge use of the IBM Q for this work. The views expressed are those of the authors and do not reflect the official policy or position of IBM or the IBM Q team. 

\bibliography{main.bib}

\end{document}